\pdfoutput=1

\documentclass[11pt]{article}

\usepackage[preprint]{coling}

\usepackage{times}
\usepackage{latexsym}

\usepackage[T1]{fontenc}

\usepackage[utf8]{inputenc}

\usepackage{microtype}

\usepackage{inconsolata}

\usepackage{graphicx}

\usepackage[most]{tcolorbox}

\newtcolorbox{contributionbox}[1]{
    enhanced,
    breakable,
    width=\linewidth,
    colback=black!2,
    colframe=black!35,
    coltitle=black,
    colbacktitle=black!6,
    boxrule=0.4pt,
    arc=1mm,
    left=5pt,
    right=5pt,
    top=4pt,
    bottom=4pt,
    before skip=5pt,
    after skip=5pt,
    fonttitle=\bfseries,
    title={#1}
}

\usepackage{graphicx}
\usepackage{amsmath}
\usepackage{amssymb}
\usepackage{booktabs}
\usepackage{tabularx}
\usepackage{array}
\usepackage{subcaption}
\usepackage{xcolor}
\usepackage{colortbl}
\usepackage{algorithm}
\usepackage{algpseudocode}
\usepackage{listings}

\definecolor{ScratchyCodeBlue}{HTML}{315D82}
\definecolor{ScratchyCodeTeal}{HTML}{2F6F6D}
\definecolor{ScratchyCodePurple}{HTML}{76558F}
\definecolor{ScratchyCodeGreen}{HTML}{3B7A57}

\lstdefinelanguage{EasyCrypt}{
  sensitive=true,
  morekeywords=[1]{module,var,proc,while,if,else,return,equiv,lemma,proof,qed},
  morekeywords=[2]{true,false,inline,wp,auto},
  morekeywords=[3]{int,list,PIR,PIR_secure1},
  morecomment=[l]{//},
  morecomment=[s]{(*}{*)},
  morestring=[b]"
}

\lstdefinestyle{scratchy-easycrypt}{
  language=EasyCrypt,
  basicstyle=\ttfamily\scriptsize,
  keywordstyle=[1]\color{ScratchyCodeBlue}\bfseries,
  keywordstyle=[2]\color{ScratchyCodePurple}\bfseries,
  keywordstyle=[3]\color{ScratchyCodeTeal},
  commentstyle=\color{ScratchyCodeGreen}\itshape,
  stringstyle=\color{ScratchyCodePurple},
  identifierstyle=\color{black},
  columns=fullflexible,
  keepspaces=true,
  showstringspaces=false,
  breaklines=true,
  tabsize=2,
  frame=TB,
  framerule=0.45pt,
  rulesep=1.2pt,
  rulecolor=\color{black!75},
  rulesepcolor=\color{black!75},
  framesep=3pt,
  aboveskip=5pt,
  belowskip=5pt,
  captionpos=b
}

\title{Scratchy: Visual-Scratchpad Multimodal Reasoning for Cryptographic Proof Generation in EasyCrypt}

\author{
  \textbf{Yupeng Ren\textsuperscript{1,2}},
  \textbf{Zhaoxuan Li\textsuperscript{1,2}},
  \textbf{Rui Zhang\textsuperscript{1,2}}
  \\
  \textsuperscript{1}State Key Laboratory of Cyberspace Security Defense,
  \\
  Institute of Information Engineering, Chinese Academy of Sciences,
  \\
  Beijing 100085, China
  \\
  \textsuperscript{2}School of Cyber Security,
  University of Chinese Academy of Sciences,
  Beijing 100049, China
  \\
  \small{\texttt{renyupeng24@mails.ucas.ac.cn}}
  \\
  \small{\texttt{\{lizhaoxuan,zhangrui\}@iie.ac.cn}}
}

\begin{document}

\maketitle
\begin{abstract}



Large language models (LLMs) have recently made substantial progress in formal proof generation, yet presenting distinctive challenges in cryptographic area. 
Computational security arguments posit that a valid proof must coordinate probability, adversarial games, invariants, assumptions and bounds, which can be provided by a machine-checked framework named EasyCrypt. 
Although all objects may appear in available context, LLMs still struggle because proof-theoretic dependencies are typically implicit in a linear representation and distributed across multiple programs.
So, this paper presents \textbf{Scratchy}, a visual-scratchpad approach that exposes these dependencies for multimodal generation.  
Given the natural-language security description, with formal context and target propositions, the proof objects can be normalized into a typed proof-relation graph.
Then a structure-preserving visual compiler transforms the graph into the formula-rich visual proof state that guides a multimodal model in generating the EasyCrypt proof.
Also, the \textbf{Scratchy-eval}, a 114-task dataset derived from reliable official EasyCrypt files, has been introduced. 
It contains 64 security-form proof generations and 50 multiple-choice knowledge tests. 
After a series of evaluations, covering semantic grounding, relational invariants, and game reductions, classical LLMs like GPT-5.6-Sol and Claude-Opus-5 have gained a clear advantage from Scratchy's structured visual proof states.
This contrast suggests that explicit proof structure can make the improvement and multimodal proof-state representation as a promising direction for computer-aided cryptography.


\end{abstract}

\section{Introduction}
\label{sec:introduction}


With the rapid development of Large Language Models (LLMs), formal proof generation has become an important task for machine reasoning. Recently, researchers have expanded formal supervision through synthetic natural language, including Lean corpora, self-play that alternates conjecturing and proving, and verifier-guided repair of generated proofs \cite{ying2024leanworkbook,dong2025stp,ospanov2025apollo}. These advances show that formal libraries and prover feedback can effectively support proof generation. However, the dominant setting remains mathematical theorem proving, and the interface between the model and the prover is still primarily composed of serialized theorem statements, proof code, candidate premises, and diagnostic messages.

\begin{figure}[t]
  \includegraphics[width=\columnwidth]{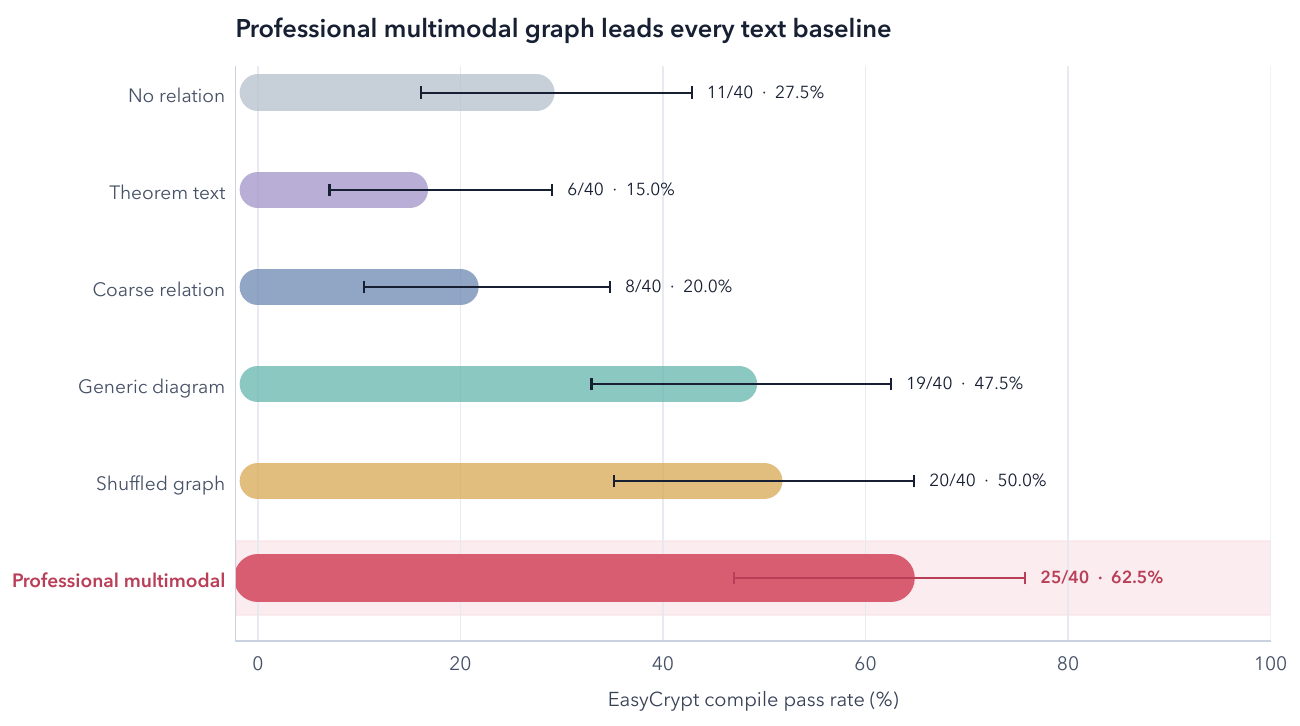}
  \caption{Results of the API-based EasyCrypt proof-generation Observation across six representation conditions. The professional multimodal graph achieves the highest compile pass rate (62.5\%, 25/40), outperforming all baselines.}
  \label{fig:experiments}
\end{figure}

\begin{table*}[t]
  \centering
  \caption{Notation expolited in the Scratchy pipeline and the motivating Observation.}
  \label{tab:scratchy-notation}
  \scriptsize
  \setlength{\tabcolsep}{3.5pt}
  \setlength{\aboverulesep}{0pt}
  \setlength{\belowrulesep}{0pt}
  \renewcommand{\arraystretch}{1.00}

  \definecolor{ScratchyBlue}{HTML}{315D82}
  \definecolor{ScratchyTeal}{HTML}{2F6F6D}
  \newcommand{\notationstrut}{\rule[-0.85ex]{0pt}{3.1ex}}
  \newcommand{\sysym}[1]{\notationstrut{\small\textcolor{ScratchyBlue}{\smash{\ensuremath{#1}}}}}
  \newcommand{\obssym}[1]{\notationstrut{\small\textcolor{ScratchyTeal}{\smash{\ensuremath{#1}}}}}

  \begin{minipage}[t]{0.488\textwidth}
    \centering
    \textbf{(a) Scratchy proof objects and verification}\par
    \begin{tabularx}{\linewidth}{@{}l >{\raggedright\arraybackslash}X@{}}
      \toprule
      \rowcolor{ScratchyBlue!9}
      \rule{0pt}{2.5ex}
      \textcolor{ScratchyBlue}{\textbf{Symbol}} &
      \textcolor{ScratchyBlue}{\textbf{Meaning}} \\
      \midrule
      \sysym{\mathcal{X},x} & \mbox{Security setting and task.} \\
      \sysym{\mathcal{C},\Gamma} & \mbox{Local context and environment.} \\
      \sysym{\phi} & \mbox{Target obligation.} \\
      \sysym{q} & \mbox{Structured Natural-Language Proof Sketch.} \\
      \sysym{\mathcal{G}=(\mathcal{V},\mathcal{E},\tau,\lambda)} & \mbox{Typed relation graph.} \\
      \sysym{\mathcal{V},\mathcal{E}} & \mbox{Graph nodes and edges.} \\
      \sysym{\tau,\lambda} & \mbox{Type and label maps.} \\
      \sysym{\mathcal{R}} & \mbox{Visual compiler.} \\
      \sysym{I_{\mathcal{G}}=\mathcal{R}(\mathcal{G})} & \mbox{Visual proof state.} \\
      \sysym{G_i,G_L,G_R} & \mbox{Hybrid and paired games.} \\
      \sysym{v_i^L,v_i^R} & \mbox{Aligned state values.} \\
      \sysym{r_i} & \mbox{Transition cue.} \\
      \sysym{\Delta_{ij}} & \mbox{Game distance.} \\
      \sysym{\epsilon_i} & \mbox{Game-hop bound.} \\
      \sysym{\pi,\widehat\pi} & \mbox{Proof fragments.} \\
      \sysym{p_\theta,\theta} & \mbox{Generator and parameters.} \\
      \sysym{\mathcal{S}[\widehat\pi]} & \mbox{Reconstructed source.} \\
      \sysym{\mathsf{V}_{\mathrm{EC}}} & \mbox{EasyCrypt verifier.} \\
      \sysym{\top,\bot} & \mbox{Verification outcomes.} \\
      \sysym{\Pr[G_i\Rightarrow1],\mathcal{I}} & \mbox{Probability and invariant.} \\
      \bottomrule
    \end{tabularx}
  \end{minipage}
  \hspace{0.012\textwidth}
  \begin{minipage}[t]{0.488\textwidth}
    \centering
    \textbf{(b) Observation metrics and routing}\par
    \begin{tabularx}{\linewidth}{@{}l >{\raggedright\arraybackslash}X@{}}
      \toprule
      \rowcolor{ScratchyTeal!9}
      \rule{0pt}{2.5ex}
      \textcolor{ScratchyTeal}{\textbf{Symbol}} &
      \textcolor{ScratchyTeal}{\textbf{Meaning}} \\
      \midrule
      \obssym{\mathcal{T},T,t} & \mbox{Task set, size, and index.} \\
      \obssym{\mathcal{M},M,m} & \mbox{Model set, size, and index.} \\
      \obssym{\mathcal{M}_{\mathrm{API}}} & \mbox{Hosted-model pool.} \\
      \obssym{\mathcal{K},K,c} & \mbox{Condition set, size, and index.} \\
      \obssym{c_{\mathrm{NR}},c_{\mathrm{TH}},c_{\mathrm{CR}}} & \mbox{Text conditions.} \\
      \obssym{c_{\mathrm{GD}},c_{\mathrm{SG}},c_{\mathrm{PG}}} & \mbox{Graph conditions.} \\
      \obssym{\mathcal{K}_{\mathrm{text}}} & \mbox{Text-condition set.} \\
      \obssym{\mathcal{K}_{\mathrm{vctl}}} & \mbox{Visual-control set.} \\
      \obssym{y_{mtc}} & \mbox{Compile outcome.} \\
      \obssym{S_{mc}=\sum_t y_{mtc}} & \mbox{Compile count.} \\
      \obssym{\widehat p_c} & \mbox{Pooled pass rate.} \\
      \obssym{[L_c,U_c]} & \mbox{Wilson interval.} \\
      \obssym{c_m^{\mathrm{text}},c_m^{\mathrm{vctl}}} & \mbox{Best text/visual controls.} \\
      \obssym{\delta_m^{\mathrm{text}},\delta_m^{\mathrm{vctl}}} & \mbox{Professional-condition gains.} \\
      \obssym{z_{mt}} & \mbox{Outcome class.} \\
      \obssym{\ell} & \mbox{Language-layer index.} \\
      \obssym{H_\ell^{(c)},\Delta H_\ell} & \mbox{Routing entropy/change.} \\
      \obssym{U_\ell^{(c)},\Delta U_\ell} & \mbox{Expert utilization/change.} \\
      \obssym{D_{\mathrm{JS}}} & \mbox{Routing divergence.} \\
      \obssym{N_{\mathrm{img}},A_{\mathrm{vis}}} & \mbox{Image tokens/activation.} \\
      \bottomrule
    \end{tabularx}
  \end{minipage}
\end{table*}


Cryptographic proofs have a different structure. A computational security argument must coordinate probabilistic programs, games, adversarial capabilities, invariants, security assumptions, and concrete advantage bounds. Correctness depends not only on whether these objects are available, but also on the direction and order required by the security reduction. Foundational work on machine-checked cryptography developed CertiCrypt for code-based game proofs and subsequently introduced EasyCrypt to elaborate machine-checkable security proofs from game sketches \cite{barthe2009formalcertification,barthe2011easycrypt}. These systems established that program transformations, relational judgments, and concrete security bounds can be rigorously checked within a formal framework. Researchers further proposed EasyCrypt proof workflow and demonstrated representative applications to cryptographic standards and post-quantum constructions \cite{barthe2012,machine-checked-standards,easypqc}.


Despite the relational of cryptographic proofs, recent proving pipelines still use textual or symbolic sequences as their primary interface. They construct natural-language with formal corpora to improve training, search, and feedback~\cite{ying2024leanworkbook,dong2025stp,wang2025treepremise,ospanov2025apollo}. However, these works do not present the distributed relationships within a proof to the generator as a unified visual state.
A linear interface preserves symbolic content, but it does not naturally expose object types. Consequently, the relationships between the direction of a game hop and multiple dependency paths running from local obligations to the security goal cannot be simultaneously represented.


Multimodal images provide a representation channel that remains underexplored. Researchers have begun to treat images as intermediate reasoning states. Visual Sketchpad allows a model to draw visual marks, GraphVis preserves graph structure through the visual modality, and MIRA evaluates problems that require intermediate visual cues \cite{hu2024visualsketchpad,deng2024graphvis,zhou2026mira}. In an adjacent formalization setting, LeanEuclid studies diagram-dependent Euclidean geometry with GPT-4V, but targets mathematical autoformalization rather than machine-checked cryptographic security proofs \cite{murphy2024autoeuclid}. 
Thus, whether task-specific images can improve machine-checkable security-proof generation, and which visual proof state is most effective, are still worth diving into.


A preliminary observation indicates that task-specific, structured proof diagrams can improve EasyCrypt proof generation across multimodal models, whereas generic or structurally corrupted diagrams provide weaker benefits. This finding suggests that the gain depends not only on the visual modality itself but also on the faithful organization of proof objects and their relations, motivating the design of Scratchy.


To sum up, this paper proposes \textbf{Scratchy}. While the security attributes are received, within local formal context, target obligation, and proof guidance, the distributed objects can be normalized into a typed proof-relation graph. Its nodes distinguish threat models, program facts, proof obligations, rules, and security goals. At the same time, the constraint, transformation, and discharge relations are supported by directed edges encoding.
The graph specifies the semantic structure that a visual representation must preserve, turning the search for an effective image into a constrained proof-state design problem.


As shown in Section.~\ref{sec:methodology}, a structure-preserving visual compiler renders the relation graph as a formula-rich visual scratchpad. Guided by different states, the multimodal model performs program unfolding, state alignment, rule application, and goal discharge to generate an EasyCrypt proof fragment. The fragment then enters an EasyCrypt verification loop for parsing and type checking, obligation discharge, and target closure.


For systematic evaluation, this paper further introduces \textbf{Scratchy-eval}. The benchmark derives 114 tasks from 26 source groups and 37 pinned official EasyCrypt files. It contains 64 free-form proof generation tasks and 50 multiple-choice tasks on semantic recognition. The tasks are evaluated under six paired representation conditions. Compilation under EasyCrypt is the primary criterion for generation.

The main contributions of this work are as follows:

\begin{contributionbox}{Empirical Finding}
This paper provides empirical evidence that structured visual proof states can improve EasyCrypt security-proof generation and that relation fidelity matters beyond image presence alone.
\end{contributionbox}

\begin{contributionbox}{Scratchy}
This paper introduces Scratchy, which treats visual-representation selection as a constrained proof-state design problem and realizes it through typed proof-relation graph, structure-preserving visual compiler, and an EasyCrypt verification loop.
\end{contributionbox}

\begin{contributionbox}{Scratchy-eval}
This paper contributes Scratchy-eval, a benchmark with 114 tasks and six paired conditions for jointly evaluating proof generation, static validity, and semantic recognition.
\end{contributionbox}


\section{Related work}
\label{sec:related-work}

\subsection{Machine-Checked Cryptographic Proofs}


Machine-checked cryptography regards security arguments as formally representable and verifiable objects, which has led to the development of CertiCrypt and EasyCrypt \cite{barthe2009formalcertification,barthe2011easycrypt}. Subsequently,  studies have further developed dedicated cryptographic-proof workflows and extensions such as EasyPQC \cite{barthe2012,machine-checked-standards,easypqc}. With program transformations, relational judgments, and probabilistic bounds, these tools provide strong support for formal cryptographic proofs and offer a correctness foundation for the generating results.


\subsection{Language Models for Cryptographic Formalization}

CrypFormBench evaluates language models on interpretation, generation, completion, transformation, and correction across cryptographic formal-verifier languages \cite{li2026crypformbench}. ShannonProver more directly targets EasyCrypt, constructing proof scripts from a supplied security model and lemma-level decomposition \cite{ma2026shannonprover}. Differently, this paper pays attention to another complementary bottleneck, which is compiling distributed formal context into a structured visual state and studies whether image presence and relation fidelity affect machine-checkable proof generation.


\subsection{Language Models for Formal Proving}

Recently, researcher expands supervision and evaluation through synthetic Lean corpora and difficult multilingual theorem benchmarks \cite{ying2024leanworkbook,tsoukalas2024putnambench}. Proof search has been improved through conjecturer-prover self-play \cite{dong2025stp}, proof-assistant-guided reinforcement learning and tree search \cite{xin2025deepseekprover}, interleaved informal reasoning and formal tactics \cite{lin2025leanstar}, scaffolded synthesis with self-correction \cite{lin2026goedelprover}, and structure-aware premise selection \cite{wang2025treepremise}. Verifier-integrated systems further repair Lean proofs from compiler feedback with Isabelle theorem proving \cite{ospanov2025apollo,lin2024fvel}. Although these methods improve data, search, and feedback, their model-facing states remain predominantly textual or symbolic and mainly target mathematical theorems or general program verification.

\begin{figure*}[t]
  \centering
  \newlength{\scratchyobservationpanelheight}
  \setlength{\scratchyobservationpanelheight}{0.267\textwidth}
  \begin{minipage}[t]{0.488\textwidth}
    \vspace{0pt}
    \centering
    \includegraphics[height=\scratchyobservationpanelheight]{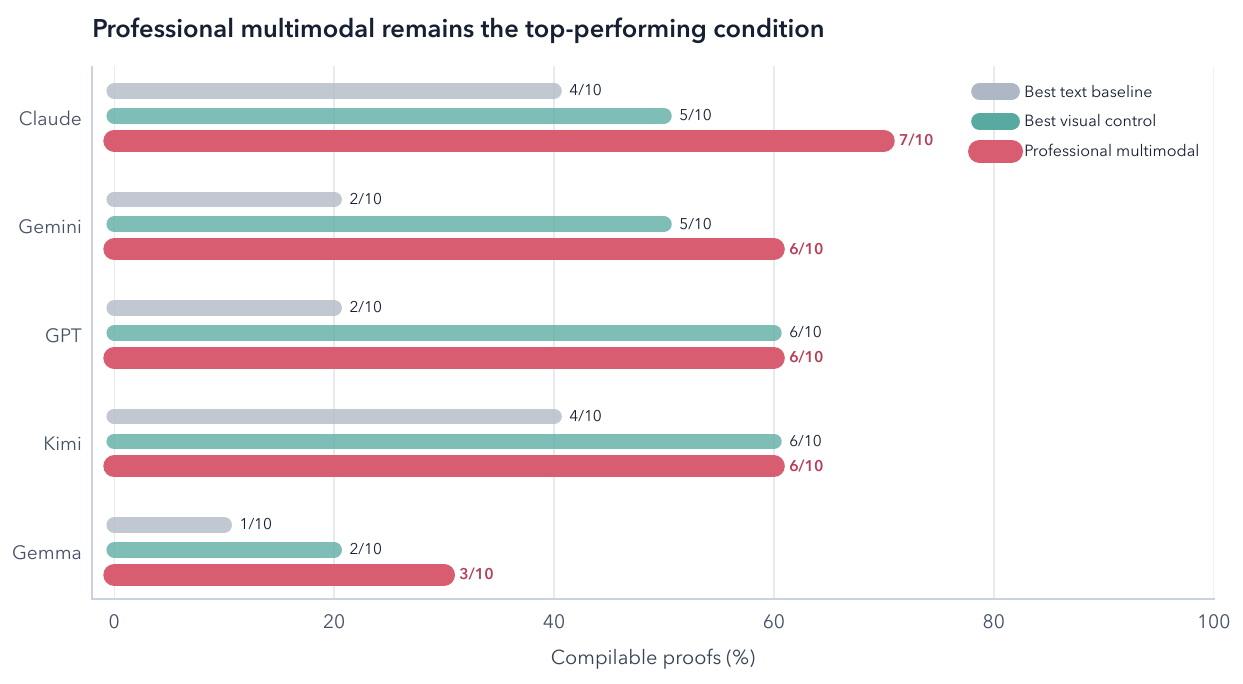}
    \par\vspace{1pt}
    \footnotesize\textbf{(a)} Model-wise comparison
  \end{minipage}
  \hfill
  \begin{minipage}[t]{0.488\textwidth}
    \vspace{0pt}
    \centering
    \includegraphics[height=\scratchyobservationpanelheight]{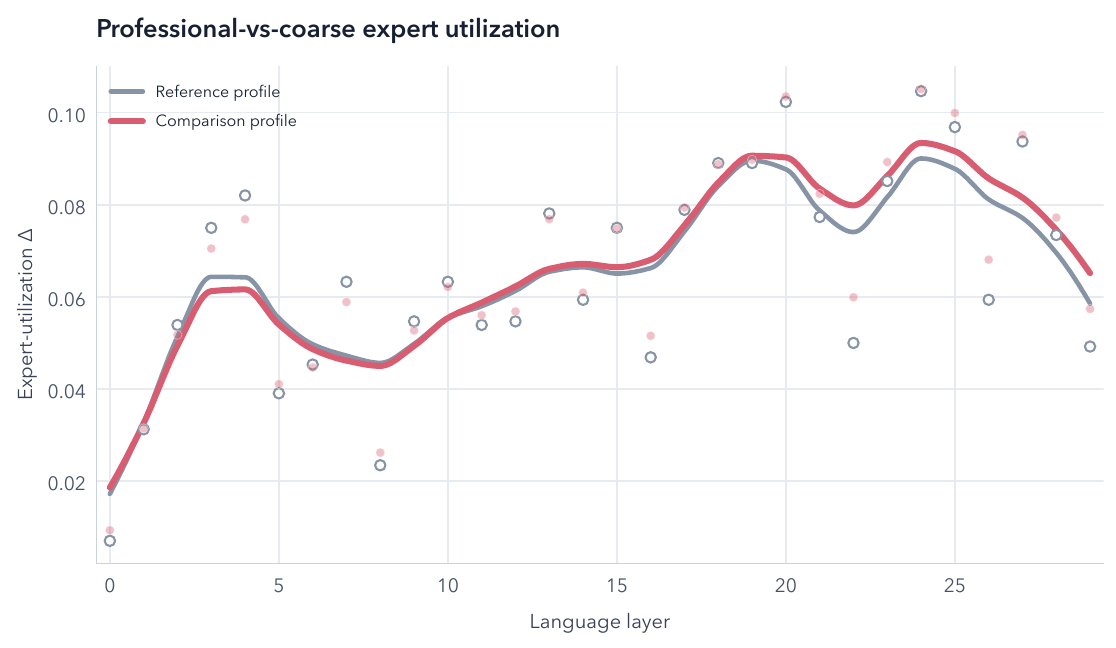}
    \par\vspace{1pt}
    \footnotesize\textbf{(b)} Expert-utilization profiles (Gemma-4-26B-A4B)
  \end{minipage}
  \caption{Complementary views of the Observation. The gray curve is the observed reference profile, and the rose curve is a comparison profile anchored to it. Models involve Claude-opus-5, Gemini-3.7-flash, Kimi-k3, GPT-5.6-sol and Gemma-4-26B-A4B.}
  \label{fig:observation-panels}
\end{figure*}


\subsection{Visual Intermediate Reasoning}

Multimodal reasoning increasingly treats images as intermediate computational states. Visualization-of-Thought, Vision-Augmented Prompting, Visual Sketchpad, and CogCoM respectively render reasoning traces, synthesize and refine visual cues, reuse drawn marks, or expose traceable visual manipulations \cite{wu2024visualization,xiao2024visionaugmented,hu2024visualsketchpad,qi2025cogcom}. GraphVis and GITA further show that visual encodings can preserve graph connectivity and that graph layout affects reasoning \cite{deng2024graphvis,wei2024gita}. Visual CoT and MIRA evaluate localized or indispensable intermediate visual evidence, while LeanEuclid studies diagram-dependent mathematical autoformalization \cite{shao2024visualcot,zhou2026mira,murphy2024autoeuclid}. However, it is still a challenge for these approaches compiles to type cryptographic proof relations into a formula-rich image whose generated proof is judged by EasyCrypt. Continuously, \textbf{Scratchy} treats relation fidelity as the central visual-design variable, which provides a good feasible solution.


\section{Observation}
\label{sec:observation}


Instructively, this paper examines a narrower question, \textbf{whether the representation of available proof relations can affect EasyCrypt proof generation with LLMs.}


This observation uses ten paired EasyCrypt obligations (Table~\ref{tab:observation-sample-sources}). Four hosted multimodal models form the behavioral summary, while a frozen local Gemma-4-26B-A4B supplies an internal diagnostic. The six conditions include no-relation, theorem-only, and coarse-relation text; a generic diagram and a shuffled-edge graph as visual controls; and a complete professional graph (Scratchy). Within a paired run, reference fragments are hidden from the model, and a generated fragment is counted as successful only when it compiles under the pinned EasyCrypt environment.

\begin{table}[t]
  \centering
  \caption{Sources of the ten paired Observation samples. All files are drawn from the pinned official EasyCrypt repository snapshot~\cite{easycrypt2026repository}.}
  \label{tab:observation-sample-sources}
  \footnotesize
  \setlength{\tabcolsep}{2.4pt}
  \renewcommand{\arraystretch}{1.04}
  \begin{tabular}{@{}ccll@{}}
    \toprule
    \textbf{ID} & \textbf{Focus} & \textbf{Proof object} & \textbf{Official source} \\
    \midrule
    01 & Semantic & Well-foundedness & \texttt{WF-examp.ec} \\
    02 & Semantic & Dice point mass & \texttt{Dice4\_6.ec} \\
    03 & Semantic & While termination & \texttt{WhileSampling.ec} \\
    04 & Relational & Async invariant & \texttt{async-while.ec} \\
    05 & Relational & Bad-event split & \texttt{FundamentalLemma.ec} \\
    06 & Reduction & Pedersen hiding & \texttt{Pedersen.ec} \\
    07 & Reduction & Schnorr soundness & \texttt{SchnorrPK.ec} \\
    08 & Reduction & ElGamal--DDH hop & \texttt{elgamal.ec} \\
    09 & Relational & PIR coupling & \texttt{PIR.ec} \\
    10 & Reduction & Stateful PRG bound & \texttt{PRG.ec} \\
    \bottomrule
  \end{tabular}
\end{table}


The aggregate API summary gives the professional graph $25/40$ compilable fragments ($62.5\%$), compared with $20/40$ ($50.0\%$) for the strongest visual control and $11/40$ ($27.5\%$) for the strongest text condition (Fig.~\ref{fig:experiments}). The model-wise view shows that professional graph is consistently above text condition, whereas its additional margin over the best visual control is smaller and varies by model (Fig.~\ref{fig:observation-panels}(a)). The contrast suggests that visual organization may provide a broad benefit, while faithful encoding of proof relations can supply a narrower, dependent contribution.


The local model provides a complementary signal about whether the visual input is computationally engaged. This paper examines the layerwise difference $\Delta U_\ell=U_\ell^{(c_{\mathrm{PG}})}-U_\ell^{(c_{\mathrm{CR}})}$ between the professional graph and coarse-relation text. The recorded reference profile and the compared profile remain close and positive across all 30 language layers (Fig.~\ref{fig:observation-panels}(b)). The trajectory changes across depth rather than appearing as a single constant offset, indicating that the representation condition is associated with a structured change in expert usage.

\begin{figure*}[t]
    \centering
    \includegraphics[width=\textwidth]{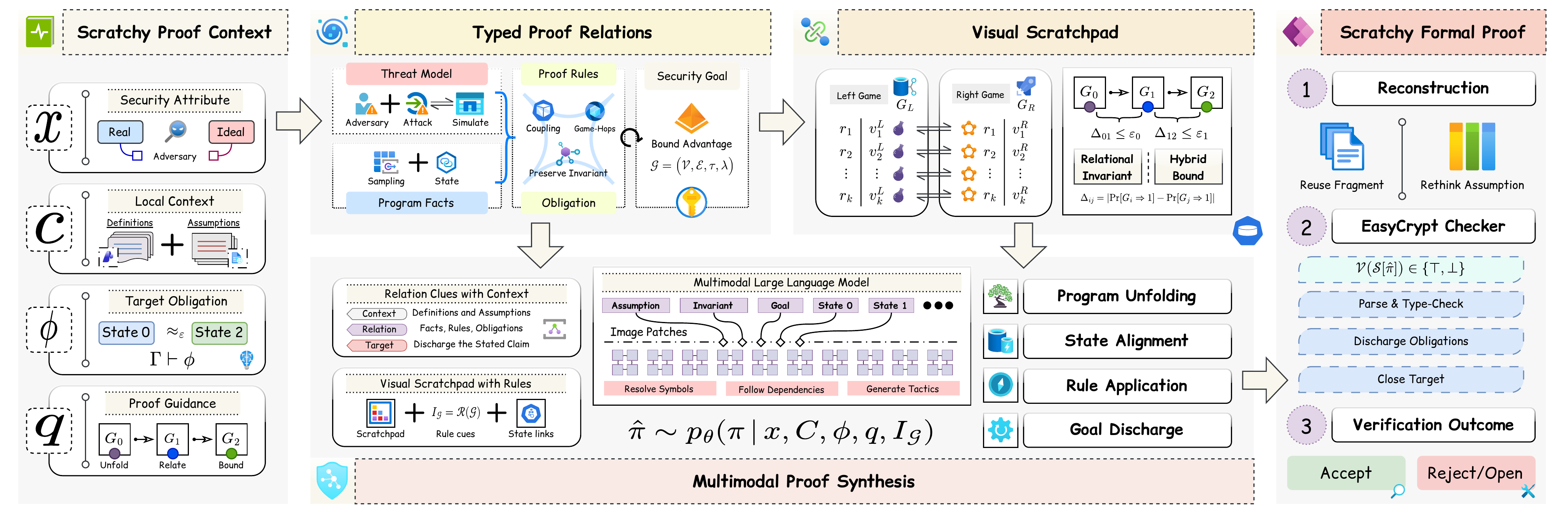}
    \caption{The Scratchy pipeline. Structured proof context and an SNPS are deterministically compiled into a typed proof-relation graph and a formula-rich visual scratchpad. A multimodal model follows the explicit dependencies to synthesize an EasyCrypt proof fragment, which is reconstructed and checked for acceptance or an open outcome.}
    \label{fig:methodology}
\end{figure*}


Therefore, the Observation shows that the representation of proof relations is itself an important part of the generation process. The performance differences across conditions indicate that structured visual states can help coordinate distributed proof objects, while the layerwise changes in expert utilization show that the model responds to this organization across network depth. These findings motivate moving beyond manually selected diagrams toward a systematic mechanism for constructing visual proof states. Accordingly, this paper introduces \textbf{Scratchy} to address these challenges, which organizes security assumptions, program states, proof obligations, and reduction goals into a structured visual interface for multimodal EasyCrypt proof generation.



\section{Methodology}
\label{sec:methodology}


In \textbf{Scratchy}, a \textbf{Structured Natural-Language Proof Sketch (SNPS)} is taken as the input to visual composition. The SNPS expresses proof objects and directed relations through fixed fields, allowing relation extraction and image construction to proceed deterministically. And LLMs enter after the visual proof state has been constructed, where it generates the EasyCrypt fragment. Fig.~\ref{fig:methodology} presents the full pipeline, and Table~\ref{tab:scratchy-notation} defines the notation.

\begin{figure*}[t]
  \centering
  \includegraphics[width=\textwidth]{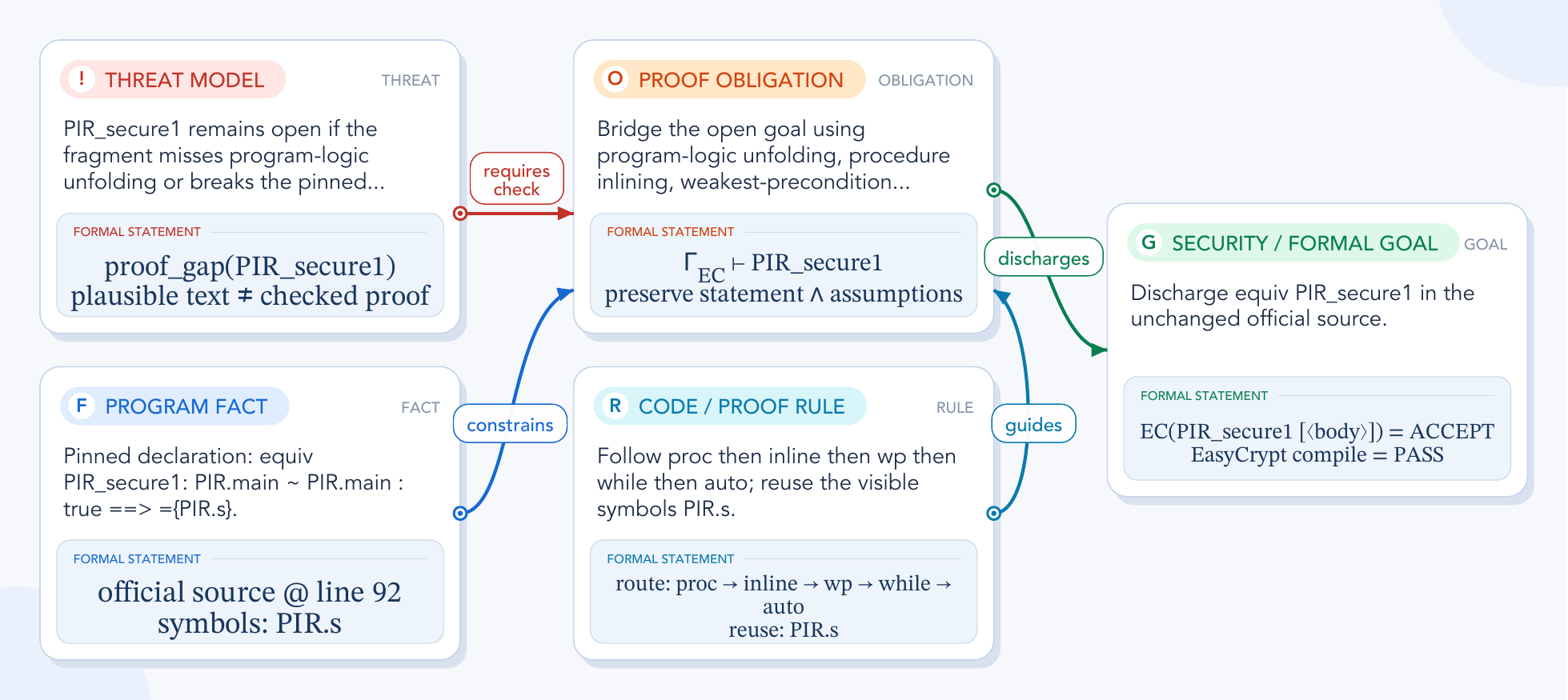}
  \caption{Professional visual proof state for a PIR relational obligation. The graph organizes source facts, the target, a proof-rule route, and the machine-checkable goal into a directed path.}
  \label{fig:snps-pir-professional-example}
\end{figure*}


\subsection{Unified Input and Output}


The unified input consists of the security setting $\mathcal{X}$, task description $x$, local EasyCrypt corpus $\mathcal{C}$, formal environment $\Gamma$, target obligation $\phi$, and SNPS $q$. The SNPS contains six fields, including \textbf{\textsc{Threat Model}, \textsc{Program Facts}, \textsc{Proof Obligations}, \textsc{Proof Rules}, \textsc{Security Goal}, and \textsc{Relations}.} The first five fields declare proof objects, \textsc{Relations} declares the source, target, and meaning of directed dependency. Identifiers and formulas in $q$ are written exactly as they occur in $\mathcal{C}$, $\Gamma$, or $\phi$.


The same interface is used throughout the pipeline:
\begin{equation}
  (\mathcal{X},x,\mathcal{C},\Gamma,\phi,q)
  \longrightarrow \mathcal{G}
  \longrightarrow I_{\mathcal{G}},
  \label{eq:scratchy-interface-visual}
\end{equation}
\begin{equation}
  I_{\mathcal{G}}
  \longrightarrow \widehat{\pi}
  \longrightarrow
  \mathsf{V}_{\mathrm{EC}}\bigl(\mathcal{S}[\widehat{\pi}]\bigr).
  \label{eq:scratchy-interface-proof}
\end{equation}
Each output is therefore the direct input to the next operation, preserving a traceable route from the SNPS to the verification outcome.


\subsection{SNPS-to-Graph Compilation}


The first key step compiles the SNPS into the typed proof-relation graph
\begin{equation}
  \mathcal{G}=(\mathcal{V},\mathcal{E},\tau,\lambda).
  \label{eq:snps-proof-relation-graph}
\end{equation}
Each SNPS item becomes a node in $\mathcal{V}$. Its field uniquely determines the node type recorded by $\tau$, while $\lambda$ retains its natural-language description, EasyCrypt identifier, and formula. Each entry in \textsc{Relations} becomes a directed edge in $\mathcal{E}$. The compiler consequently performs parsing, matching, and normalization without inferring relations absent from the SNPS.


The two deterministic procedures are given in
Algorithms~\ref{alg:snps-to-graph} and~\ref{alg:graph-to-visual}.

\begin{algorithm}[t]
  \captionsetup{font=footnotesize,labelfont=bf}
  \caption{Deterministic SNPS-to-Graph Compilation}
  \label{alg:snps-to-graph}
  \small
  \begin{algorithmic}[1]
    \Require $q$ grounded in $(\mathcal{X},x,\mathcal{C},\Gamma,\phi)$
    \Ensure $\mathcal{G}=(\mathcal{V},\mathcal{E},\tau,\lambda)$
    \State Parse the six ordered fields of $q$
    \ForAll{proof objects declared in $q$}
      \State Add a node to $\mathcal{V}$; assign $\tau$ and $\lambda$
    \EndFor
    \ForAll{source-to-target relations in $q$}
      \State Add the corresponding directed edge to $\mathcal{E}$
    \EndFor
    \State Order dependency paths toward $\phi$
    \State \Return $\mathcal{G}$
  \end{algorithmic}
\end{algorithm}

\begin{algorithm}[t]
  \captionsetup{font=footnotesize,labelfont=bf}
  \caption{Structure-Preserving Visual Compilation}
  \label{alg:graph-to-visual}
  \small
  \begin{algorithmic}[1]
    \Require $\mathcal{G}=(\mathcal{V},\mathcal{E},\tau,\lambda)$
    \Ensure $I_{\mathcal{G}}$
    \State Arrange nodes by the roles recorded in $\tau$
    \State Render the text, formulas, and code recorded in $\lambda$
    \ForAll{directed edges in $\mathcal{E}$}
      \State Draw its tail, arrowhead, and relation label
    \EndFor
    \State Highlight the dependency paths ending at $\phi$
    \State \Return $I_{\mathcal{G}}=\mathcal{R}(\mathcal{G})$
  \end{algorithmic}
\end{algorithm}


There are distinct roles for five node types. A threat-model node states the security consequence addressed by the proof; program-fact nodes provide available evidence; proof-obligation nodes state intermediate claims; proof-rule nodes describe permitted transformations; and the security-goal node corresponds to $\phi$. Edges point from a justification to the claim it supports. And both proof order and dependency are recorded by the complete path.


\subsection{Graph-to-Visual Compilation}


The second key step applies visual compiler $\mathcal{R}$:
\begin{equation}
  I_{\mathcal{G}}=\mathcal{R}(\mathcal{G}).
  \label{eq:snps-visual-compilation}
\end{equation}
Structure preservation means that node roles, semantic labels, edge directions, and the path to $\phi$ remain directly readable in the output. Role-specific colors and icons encode $\tau$; formula and code cards display $\lambda$; and tail markers, arrowheads, and relation labels encode $\mathcal{E}$. a stable, robust, and professional layout is constructed.


For relational proofs, $I_{\mathcal{G}}$ places $G_L$ and $G_R$ in parallel and connects $v_i^L$, $r_i$, and $v_i^R$ at corresponding program points. For game-based proofs, it orders $G_i$ along the reduction and displays next to the relevant hop.
\begin{equation}
  \scalebox{0.88}{$\displaystyle
    \Delta_{ij}=
    \left|\Pr[G_i\!\Rightarrow\!1]-\Pr[G_j\!\Rightarrow\!1]\right|,
    \quad \Delta_{ij}\leq\epsilon_i
  $}
  \label{eq:snps-game-distance}
\end{equation}

The relational invariant $\mathcal{I}$ is placed beside the state correspondence it constrains.


For example, Fig.~\ref{fig:snps-pir-professional-example} illustrates the resulting visual proof state with a representative PIR obligation. The corresponding EasyCrypt excerpt is shown separately in
Listing~\ref{lst:snps-pir-code-example}.

\begin{lstlisting}[
  style=scratchy-easycrypt,
  float=t,
  caption={Abridged EasyCrypt source for the PIR example. The relational target and proof fragment instantiate the SNPS route encoded by the professional graph.},
  label={lst:snps-pir-code-example}]
module PIR = {
  var s, s' : int list
  proc main(i : int) = {
    var j <- 0;
    (s, s') <- ([], []);
    while (j < N) {
      (* sample and update s, s' *)
      j <- j + 1;
    }
  }
}.

equiv PIR_secure1:
  PIR.main ~ PIR.main : true ==> ={PIR.s}.
proof.
  proc; inline *; wp.
  while (={j, PIR.s}); auto.
qed.
\end{lstlisting}


\subsection{Proof Synthesis and Verification}


After $I_{\mathcal{G}}$ has been constructed, the multimodal generator reads both the textual context and the image:
\begin{equation}
  \widehat{\pi}\sim
  p_\theta(\pi\mid x,\mathcal{C},\phi,q,I_{\mathcal{G}}).
  \label{eq:snps-proof-generation}
\end{equation}
The graph route organizes generation into program unfolding, state alignment, rule application, and goal discharge. Text preserves exact EasyCrypt syntax, while $I_{\mathcal{G}}$ preserves the relation structure needed to produce $\widehat{\pi}$.


Finally, Scratchy inserts $\widehat{\pi}$ into its original location to form $\mathcal{S}[\widehat{\pi}]$. EasyCrypt parses and type-checks the reconstructed source, discharges its obligations, and closes the target:
\begin{equation}
  \mathsf{V}_{\mathrm{EC}}\bigl(\mathcal{S}[\widehat{\pi}]\bigr)
  \in\{\top,\bot\}.
  \label{eq:snps-easycrypt-verification}
\end{equation}
An outcome of $\top$ accepts the fragment, while $\bot$ returns the open proof state to reconstruction. Algorithms~\ref{alg:snps-to-graph} and~\ref{alg:graph-to-visual} remain fully deterministic. The LLM is used only for the transition from $I_{\mathcal{G}}$ to $\widehat{\pi}$.



\begin{table*}[t]
  \centering
  \caption{Scratchy-eval results under six representation conditions. Each numeric entry is passed/total (\%). NR: no relation knowledge; TH: theorem-only text; CR: coarse proof-relation text; GD: generic diagram; SG: shuffled-edge graph; PG: Scratchy professional graph.}
  \label{tab:scratchy-eval-main}
  \scriptsize
  \renewcommand{\arraystretch}{1.06}
  \newcommand{\modellogo}[1]{\raisebox{-0.18\height}{\includegraphics[height=1.35em]{#1}}\hspace{0.42em}}
  \begin{tabular*}{\textwidth}{@{\extracolsep{\fill}}lcccccc@{}}
    \toprule
    \textbf{Model or task set} & \textbf{NR} & \textbf{TH} & \textbf{CR} & \textbf{GD} & \textbf{SG} & \textbf{PG} \\
    \midrule
    \modellogo{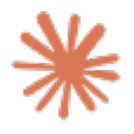}Claude Opus 5    & 0/64 (0.0) & 3/64 (4.7) & 5/64 (7.8) & 1/64 (1.6) & 1/64 (1.6) & \textbf{9/64 (14.1)} \\
    \modellogo{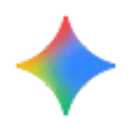}Gemini 3.7 Flash & 2/64 (3.1) & 2/64 (3.1) & 3/64 (4.7) & 9/64 (14.1) & 9/64 (14.1) & \textbf{18/64 (28.1)} \\
    \modellogo{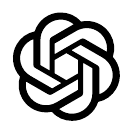}GPT-5.6-Sol     & 2/64 (3.1) & 3/64 (4.7) & 2/64 (3.1) & 4/64 (6.2) & 6/64 (9.4) & \textbf{12/64 (18.8)} \\
    \modellogo{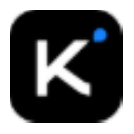}Kimi K3          & 1/64 (1.6) & 1/64 (1.6) & 1/64 (1.6) & 5/64 (7.8) & 8/64 (12.5) & \textbf{15/64 (23.4)} \\
    \midrule
    Pooled generation   & 5/256 (2.0) & 9/256 (3.5) & 11/256 (4.3) & 19/256 (7.4) & 24/256 (9.4) & \textbf{54/256 (21.1)} \\
    Pooled MCQ          & 194/200 (97.0) & 199/200 (99.5) & 198/200 (99.0) & 198/200 (99.0) & 199/200 (99.5) & \textbf{199/200 (99.5)} \\
    Overall             & 199/456 (43.6) & 208/456 (45.6) & 209/456 (45.8) & 217/456 (47.6) & 223/456 (48.9) & \textbf{253/456 (55.5)} \\
    \bottomrule
  \end{tabular*}
\end{table*}


\section{Experiments}
\label{sec:evaluation}


\subsection{Scratchy-eval}


\textbf{Scratchy-eval} is a provenance-preserving benchmark constructed from the EasyCrypt corpus~\cite{easycrypt2026repository}. It contains 114 tasks drawn from 26 protocol or formal-program sources and 37 official files (13,761 lines). 64 tasks require free-form EasyCrypt proof generation, while 50 multiple-choice tasks cover static validity and semantic recognition. 

The generation set comprises 55 lemmas, seven equivalence judgments, and two expectation-Hoare judgments. Each generation fixture preserves the imports and dependency prefix of an official target but replaces its proof. Static-validity items pair an official statement with syntactically defective alternatives, and semantic items test the formal role of local constructs. Appendix~\ref{sec:datacard} provides the Data Card and summarizes source coverage and construction.


\begin{figure}[t]
  \centering
  \includegraphics[width=\columnwidth]{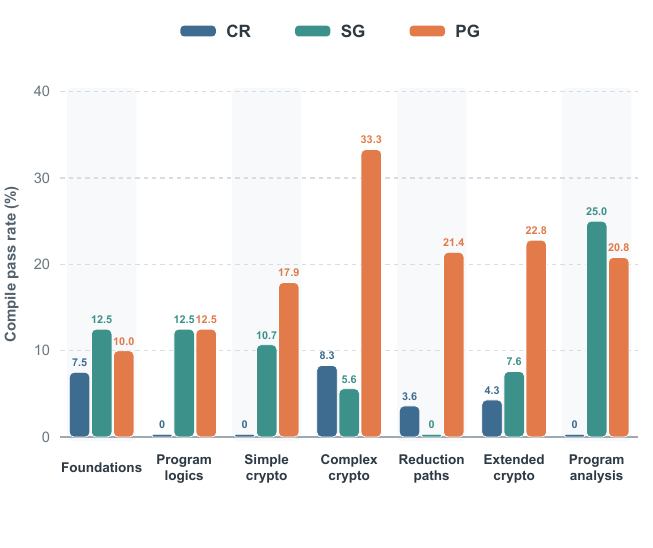}
  \caption{Generation pass rates across seven source families, pooled over four models. Each vertical group compares CR (coarse proof-relation text), SG (shuffled-edge graph), and PG (Scratchy professional graph); bar height and top labels report the percentage of fragments accepted by EasyCrypt. PG leads on the four cryptography and reduction-oriented families, SG leads on foundations and program analysis, and the two graph conditions tie on program logics.}
  \label{fig:scratchy-eval-source-family-bars}
\end{figure}


\subsection{Experimental Setup}


This paper evaluates Claude Opus 5, Gemini 3.7 Flash, GPT-5.6-Sol, and Kimi K3. Each model produces one response for every task under six paired representation conditions. The complete condition receives the local source, task-specific professional graph, and EasyCrypt decoding card; the other conditions remove source details, proof paths, identifiers, or edge structure to test which parts of the visual proof state contribute to generation.


Each model run contains $684$ units, giving 2,736 model task condition units in total. The task, response contract, and output budget remain fixed within each six-way comparison, while condition order is deterministically rotated by task. Before any model call, 164 fixtures are dynamically checked: the 64 hidden reference proofs and four isolated options for each of the 25 static-validity tasks.


The primary generation metric is \emph{compile pass rate}: a fragment must parse, type-check, discharge its obligations, and close the target in the pinned EasyCrypt environment. Exact agreement with the official fragment is only diagnostic because several valid proofs may establish the same target. Multiple-choice tasks use accuracy, with static-validity selections independently confirmed by compilation. Paired confidence intervals are obtained by bootstrap resampling the 26 protocol or formal-program sources, preserving dependencies among tasks derived from the same source.

\subsection{Main Results}


Table~\ref{tab:scratchy-eval-main} shows a consistent model-level result: PG is the strongest generation condition for all four models. Pooled over models, it compiles 54 of 256 fragments ($21.1\%$), compared with 24 ($9.4\%$) under SG and 11 ($4.3\%$) under CR. The complete representation therefore reaches $2.25\times$ the SG rate and $4.91\times$ the CR rate. PG ranges from $14.1\%$ on Claude to $28.1\%$ on Gemini, showing that the ordering is stable even though absolute proof ability differs across models.


The multiple-choice tasks remain ($97.0$--$99.5\%$), making free-form generation the principal source of discrimination. Among the 54 compiling PG outputs, only 10 exactly match the official fragment. The remaining 44 are textually different proofs accepted by EasyCrypt, which would be missed by reference-only scoring.


\subsection{Protocol and Proof-Structure Analysis}


Figure~\ref{fig:scratchy-eval-source-family-bars} locates the gain precisely. PG reaches $33.3\%$ on complex cryptography and $21.4\%$ on reduction paths. Generic visual organization can be sufficient for more local program facts, whereas the largest professional-graph advantage appears when a target depends on a protocol-specific reduction path.


At the individual-source level, PG compiles $9/20$ model--target pairs for two-server PIR, $7/16$ for UC Diffie--Hellman encryption, $5/16$ for syntactic up-to reductions, and $5/16$ for the PRG tutorial. It also obtains $3/12$ on Pedersen commitment where both CR and SG obtain zero. The broader source coverage behind these counts is summarized in Table~\ref{tab:datacard-task-protocol-map}.


The same pattern appears across formal judgment types. PG compiles $49/220$ lemma pairs ($22.3\%$), compared with $23/220$ for SG ($10.5\%$) and $10/220$ for CR ($4.5\%$). For relational \texttt{equiv} targets, PG reaches $5/28$ ($17.9\%$), whereas both SG and CR reach $1/28$ ($3.6\%$). None of the eight model--task pairs formed by the two \texttt{ehoare} targets compiles under any condition.


\begin{table}[t]
  \centering
  \caption{Paired generation comparison over 256 model--task pairs. PG: Scratchy professional graph; CR: coarse proof-relation text; SG: shuffled-edge graph; W/L/T: PG wins/losses/ties; pp: percentage points; CI: source-group bootstrap 95\% confidence interval.}
  \label{tab:scratchy-eval-paired}
  \footnotesize
  \setlength{\tabcolsep}{4.0pt}
  \renewcommand{\arraystretch}{1.06}
  \begin{tabular}{@{}cccc@{}}
    \toprule
    \textbf{Compared with} & \textbf{W/L/T} & \textbf{Gain (pp)} & \textbf{95\% CI} \\
    \midrule
    CR & 53/10/193 & +16.8 & [8.8, 24.6] \\
    SG & 39/9/208  & +11.7 & [3.5, 19.8] \\
    \bottomrule
  \end{tabular}
\end{table}


The paired comparison in Table~\ref{tab:scratchy-eval-paired} controls for both model and target. Against CR, PG turns 53 failures into passes while losing 10 previously passing pairs; against SG, the corresponding counts are 39 and nine. Both source-group intervals remain above zero. Moreover, 36 of the 54 PG successes are not reproduced by any of the five ablations for the same model and target. Fourteen are shared with a visual control, two with a text condition, and two with both families. The aggregate improvement is thus composed mainly of new machine-accepted proofs, rather than repeated successes on tasks already handled by simpler representations.


\subsection{Failure and Efficiency Analysis}


Among the unsuccessful PG generation units, the largest recurring groups are parse errors (30), inapplicable rewrites reported as \textbf{nothing to rewrite} (30), incomplete proofs (19), and \texttt{by}-blocks that cannot close their goals (11). Together they account for 90 failures ($44.6\%$). The remaining bottleneck is therefore often the local instantiation of the proof route in exact EasyCrypt syntax.



\section{Discussion}
\label{sec:discussion}


\subsection{A Machine-Checked Proof Contrast}

Task 064 provides a concrete view of what changes between a successful and an unsuccessful generation. The target is the relational judgment shown in Figure~\ref{fig:snps-pir-professional-example}. Listing~\ref{lst:pir-proof-contrast} compares the official fragment with two outputs from GPT-5.6-Sol for this same target. The official fragment and the professional-graph output are accepted by EasyCrypt.


\begin{lstlisting}[
  style=scratchy-easycrypt,
  float=t,
  caption={Machine-checked contrast for the PIR target. PG denotes the professional graph and CR denotes coarse relation text. The official and PG fragments compile, whereas CR omits the loop invariant and leaves an open goal.},
  label={lst:pir-proof-contrast}]
(* Official reference: accepted *)
proc; inline *; wp.
while (={j,PIR.s}); auto.

(* GPT + PG: accepted *)
proc; inline *; wp;
while (={j} /\ ={PIR.s}).
  wp; rnd; skip; auto.
auto.

(* GPT + CR: rejected *)
by
  proc;
  inline *;
  wp;
  auto.
\end{lstlisting}


All three fragments begin with the same local route, including expose the paired procedures with \texttt{proc}, unfold calls with \texttt{inline}, and propagate postconditions with \texttt{wp}. The decisive step is the loop. The rejected output invokes autoimmediately and never states what must remain synchronized across iterations. 

In contrast, the professional-graph output aligns the random assignment with \texttt{rnd}, and discharges the loop body before closing the residual goal. Unlike the official proof, GPT produces the equivalent conjunction and an explicit body proof. Machine checking therefore identifies a valid alternative that textual exact match would mark as different.


\subsection{Implications for Scratchy}

This example reflects the aggregate pattern. The professional graph is strongest for all four models and yields its clearest gains on protocol-specific reductions and relational judgments. For Task 064, that route makes the preserved state and the loop rule visible at the point where a linear hint remains underspecified. The result supports Scratchy's central design which compiling an SNPS into a relation-faithful visual state before generation.

Many remaining failures occur when a plausible route must be instantiated as exact syntax, especially around rewrites, side goals, and program-logic tactics. This suggests that  visual compiler should continue to expose proof structure, while the decoder and verification loop should become more precise at converting that structure into locally valid EasyCrypt steps. Extending verifier-guided revision at this final boundary is likely to improve proof completion without changing the deterministic visual construction.


\section{Conclusion}
\label{sec:conclusion}


This paper presents Scratchy, a visual-scratchpad framework for machine-checkable cryptographic proof generation. Scratchy deterministically compiles a Structured Natural-Language Proof Sketch into a typed proof-relation graph and a formula-rich visual state, then guides a multimodal model to produce an EasyCrypt fragment whose correctness is decided by the formal verifier. Across Scratchy-eval, the professional graph consistently improves generation over textual and visual-control conditions, with particularly useful gains for relational and reduction-oriented proofs.


Future work will extend Scratchy-eval to additional protocols and proof systems, refine SNPS authoring and visual compilation, and use verifier feedback for targeted repair of syntax and residual obligations. A further direction is to study which visual relations are most useful for different proof families, enabling proof states that remain concise while adapting their structure to the target judgment.

\bibliography{custom}

\appendix



\section{Scratchy-eval Data Card}
\label{sec:datacard}


Scratchy-eval is designed to compare how alternative representations of proof relations affect machine-checkable EasyCrypt generation. It also includes compact diagnostic tasks for static validity and semantic recognition. The intended unit of comparison is a fixed model, task pair evaluated under the same response contract and verifier.


This dataset contains 114 tasks, including 64 free-form proof-generation targets and 50 multiple-choice diagnostics. The source material spans 26 protocol or formal-program groups. Table~\ref{tab:datacard-task-protocol-map} reports category-level coverage; task-level source pointers are retained in the accompanying metadata.

\subsection{Representative datum}


Task 003 is drawn from the official while-sampling example in the foundations family. Its source is sufficiently compact to reproduce in full. Listing~\ref{lst:datacard-official-example} includes the complete pinned file: imports, declarations, probabilistic program, target, and accepted proof. To construct the generation task, only the body of \texttt{Sample\_lossless} is replaced by a proof hole.

\begin{lstlisting}[
  style=scratchy-easycrypt,
  float=t,
  caption={Complete official EasyCrypt source underlying Scratchy-eval Task 003. The task removes the proof body of \texttt{Sample\_lossless} while preserving the surrounding file as its compilation context.},
  label={lst:datacard-official-example}]
require import Real Distr.

type t.

op sample: t distr.
axiom sample_ll: is_lossless sample.

op test: t -> bool.
axiom pr_ntest: 0%r < mu sample (predC test).

module Sample = {
  proc sample () : t = {
    var r : t;

    r <$ sample;
    while (test r) {
      r <$ sample;
    }
    return r;
  }
}.

lemma Sample_lossless: islossless Sample.sample.
proof.
proc; seq  1: true=> //.
+ by auto=> />; exact/sample_ll.
while true (if test r then 1 else 0) 1 (mu sample (predC test))=> //.
+ by move=> _ r; case: (test r).
+ move=> ih; seq  1: true=> //.
  by auto; rewrite sample_ll.
+ by auto; rewrite sample_ll.
rewrite pr_ntest=> /= z; conseq (: true ==> !test r).
+ smt().
by rnd; auto=> />.
qed.
\end{lstlisting}


The generation contract asks for the missing proof body rather than the surrounding declaration. Listing~\ref{lst:datacard-task003-output} gives the complete canonical output for Task 003. This fragment reconstructs the official source above and is accepted by EasyCrypt; a textually different response is also correct when it closes the same target under the verifier.

\begin{lstlisting}[
  style=scratchy-easycrypt,
  float=t,
  caption={Complete accepted output for Scratchy-eval Task 003. This canonical fragment is taken from the official answer record and contains exactly the code required at the proof hole.},
  label={lst:datacard-task003-output}]
proc; seq  1: true=> //.
+ by auto=> />; exact/sample_ll.
while true (if test r then 1 else 0) 1 (mu sample (predC test))=> //.
+ by move=> _ r; case: (test r).
+ move=> ih; seq  1: true=> //.
  by auto; rewrite sample_ll.
+ by auto; rewrite sample_ll.
rewrite pr_ntest=> /= z; conseq (: true ==> !test r).
+ smt().
by rnd; auto=> />.
\end{lstlisting}


\begin{table*}[t]
  \centering
  \caption{Category-level coverage of Scratchy-eval. Representative material and proof emphasis are summarized without enumerating task IDs or hidden fragments. The generation counts sum to 64; static-validity and semantic-recognition diagnostics accompany all seven families.}
  \label{tab:datacard-task-protocol-map}
  \small
  \setlength{\tabcolsep}{3.8pt}
  \renewcommand{\arraystretch}{1.16}
  \begin{tabular*}{0.98\textwidth}{@{\extracolsep{\fill}}ccccc@{}}
    \toprule
    \textbf{Source family} & \textbf{Representative material} &
    \textbf{Proof emphasis} & \textbf{Judgments} & \textbf{Gen.} \\
    \midrule
    Foundations & Recursion; distributions; sampling & Termination and probability & \texttt{lemma/equiv} & 10 \\
    Program logics & Relational while programs & Coupling and invariants & \texttt{lemma} & 2 \\
    Simple crypto & Pedersen; Schnorr; ElGamal & Correctness and basic security & \texttt{lemma} & 7 \\
    Complex crypto & PIR; stateful PRG & Relational privacy and state & \texttt{lemma/equiv} & 9 \\
    Reduction paths & Up-to; plug-and-pray & Game transitions and bounds & \texttt{lemma} & 7 \\
    Extended crypto & AE; UC/CCA; hashed ElGamal & Composed and oracle reductions & \texttt{lemma/equiv} & 23 \\
    Program analysis & Adversary bounds; Quickselect & Expectations and data invariants & \texttt{lemma/ehoare} & 6 \\
    \bottomrule
  \end{tabular*}
\end{table*}


For generation, a declaration is selected from its source and the required imports, definitions, and dependency prefix are preserved while the official proof is replaced. The hidden fragment is used for fixture preflight and analysis, but is not shown to the model. The multiple-choice portion asks either whether a local EasyCrypt form is valid or which formal role a construct serves. All answers are checked against the pinned source or compiler behavior.


\subsection{Representations and scoring}

Each task supports three textual conditions and three graph conditions. The professional graph presents typed proof objects, local formulas, directed dependencies, and a route to the target; the remaining conditions remove relation detail or alter visual structure. Generation is scored by EasyCrypt acceptance after source reconstruction, while the diagnostic tasks use answer accuracy. This separates executable proof completion from recognition of formal knowledge.


The release record retains source provenance, task type, target information, graph metadata, supported conditions, and verification artifacts. Scratchy-eval is intended for controlled representation studies over the included EasyCrypt material. Its category-level coverage is broader than the motivating Observation, while remaining a focused benchmark rather than a census of cryptographic formalization.

\end{document}